\documentclass[]{spie}  

\newcommand*\arcmin{\ensuremath{^\prime}}
\newcommand*\arcsec{\ensuremath{^{\prime\prime}}}
 
\usepackage{amsmath,amsfonts,amssymb}
\usepackage{graphicx}
\usepackage[colorlinks=true, allcolors=blue]{hyperref}

\title{Regaining the FORS: making optical ground-based transmission spectroscopy of exoplanets with VLT+FORS2 possible again}

\author[a,b]{Henri M.J. Boffin}
\author[a,b]{Elyar Sedaghati}
\author[a,c]{Guillaume Blanchard}
\author[a,d]{Oscar Gonzalez}
\author[b]{Sabine Moehler}
\author[b,e]{Neale Gibson}
\author[b]{Mario van den Ancker}
\author[a]{Jonathan Smoker}
\author[a]{Joseph Anderson}
\author[b]{Christian Hummel}
\author[b]{Danuta Dobrzycka}
\author[a]{Alain Smette}
\author[b]{Gero Rupprecht}

\affil[a]{ESO, Av. Alonso de Cordova 3107, Casilla 19001, Santiago 19, Chile}
\affil[b]{ESO, Karl-Schwarzschild-str. 2, 85748 Garching, Germany}
\affil[c]{OPA-Opticad, 9 Rue L\'eon Foucault, 77295 Mitry-Mory CEDEX, France}
\affil[d]{UK Astronomy Technology Centre, 
Royal Observatory,
Blackford Hill,
Edinburgh
EH9 3HJ, UK}
\affil[e]{Astrophysics Research Centre, School of Mathematics and Physics, Queens University Belfast, UK}
\authorinfo{Send correspondence to H.M.J.B.\\ E-mail: hboffin@eso.org}

\authorinfo{Send correspondence to H.M.J.B.\\ E-mail: hboffin@eso.org}

\begin{document} 
\maketitle

\begin{abstract}
Transmission spectroscopy {facilitates the detection} of molecules and/or clouds in the atmospheres of exoplanets. Such studies rely heavily on space-based or large ground-based {observatories}, as one needs to perform time-resolved, high signal-to-noise spectroscopy. The FORS2 instrument at ESO's Very Large Telescope is the obvious choice for {performing} such studies, and was indeed pioneering the field in 2010. After that, however, it was shown to suffer from systematic errors caused by the Longitudinal Atmospheric Dispersion Corrector (LADC). This was successfully addressed, leading to a renewed interest for this instrument as shown by the number of proposals submitted to perform transmission spectroscopy of exoplanets. We present here the context, the problem and how we solved it, as well as the recent results obtained. We finish by providing tips for an optimum strategy to do transmission spectroscopy with FORS2, in   the hope that FORS2 may become the instrument of choice for ground-based transmission spectroscopy of exoplanets.\end{abstract}


\section{Transmission Spectroscopy}
\label{sec:intro}  

Since the first discovery in 1995, we now know more than 3\,400 exoplanets and many more candidates are awaiting confirmation. Although the first exoplanets were discovered thanks to the radial-velocity technique, the biggest batch of exoplanets has been discovered through photometry: those are transiting exoplanets, i.e. those that pass between their host star and us, leading to a dimming of the star's light. In addition to the CoRoT and Kepler satellites that provided the bulk of all transiting exoplanet candidates (and, soon, TESS), there are a large number of ground-based surveys in operation or planned (such as WASP, HAT, NGTS or HAT-PI).

These discoveries have revealed an amazing variety of alien worlds, the existence of which we could not have imagined, from `hot Jupiters' to `ocean worlds' and `super-Earths', and provided in turn the necessary insight to reassess our understanding of the formation of the Solar System, which must have been formed in a much more dynamic way than previously thought. Moreover, exoplanetary science has evolved from a pure discovery endeavour to a more physical science, with the aim to characterise and understand the physical properties of exoplanets -- masses, radii, and thus densities, but also bulk composition and more recently their atmospheres. Studying the 
atmosphere not only gives us clues about the original place of a planet's formation, but it also helps to break degeneracies that remain in the models describing the internal structure of a planet (e.g., Ref.~\citenum{Valencia13}). 

\begin{figure} [ht]
   \begin{center}
   \begin{tabular}{c} 
   \includegraphics[width=17cm]{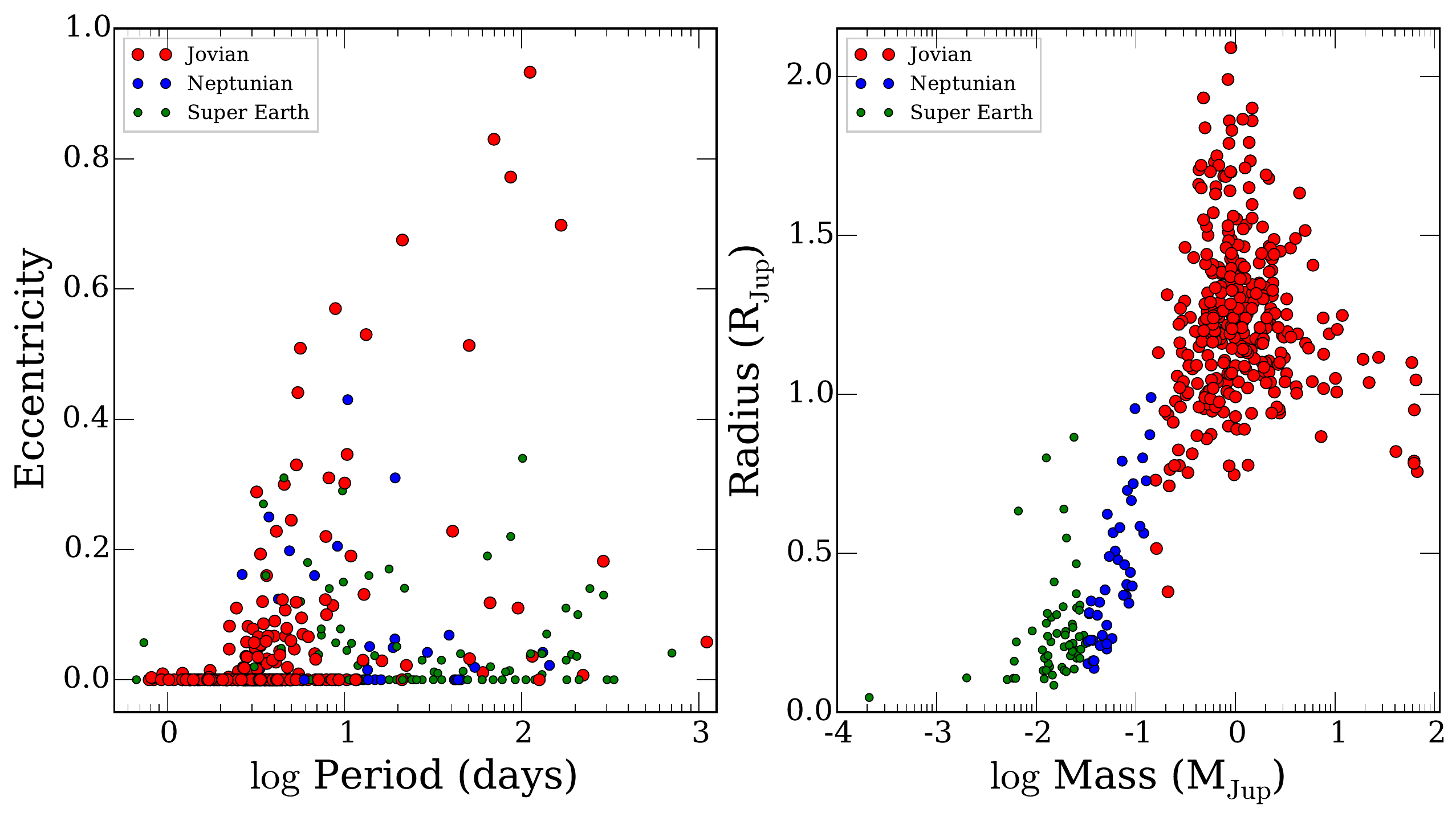} 
   \end{tabular}
   \end{center}
   \caption[example] 
   { \label{Fig:Transit} 
Parameters of known transiting exoplanets: the left panel shows the eccentricity as a function of the orbital period (in days; log scale), while the right panel shows the radius vs. mass for those where these values could be established.}
   \end{figure}

Luckily, the most numerous known exoplanets -- transiting ones -- are also those that, potentially, would allow us to get the most detailed information about planetary atmospheres. According to the {\tt exoplanet.eu} online catalogue, at the time of writing, there are 2\,619 transiting planets distributed among 1\,953 planetary systems. These planets  have orbital periods between 1.3~hours and 10 years, and radii between 0.0276 and 2.1 times the radius of Jupiter (R$_{\rm Jup}$). The distribution of some of the most important parameters are shown in Fig.~\ref{Fig:Transit}. Of course, not all of these have been characterised well enough. TEPCat\footnote{\url{http://www.astro.keele.ac.uk/jkt/tepcat/}} quotes 540 well-studied transiting planets, for which the planetary and stellar properties are well known. 

\begin{figure} [ht]
   \begin{center}
   \begin{tabular}{c} 
   \includegraphics[width=17cm]{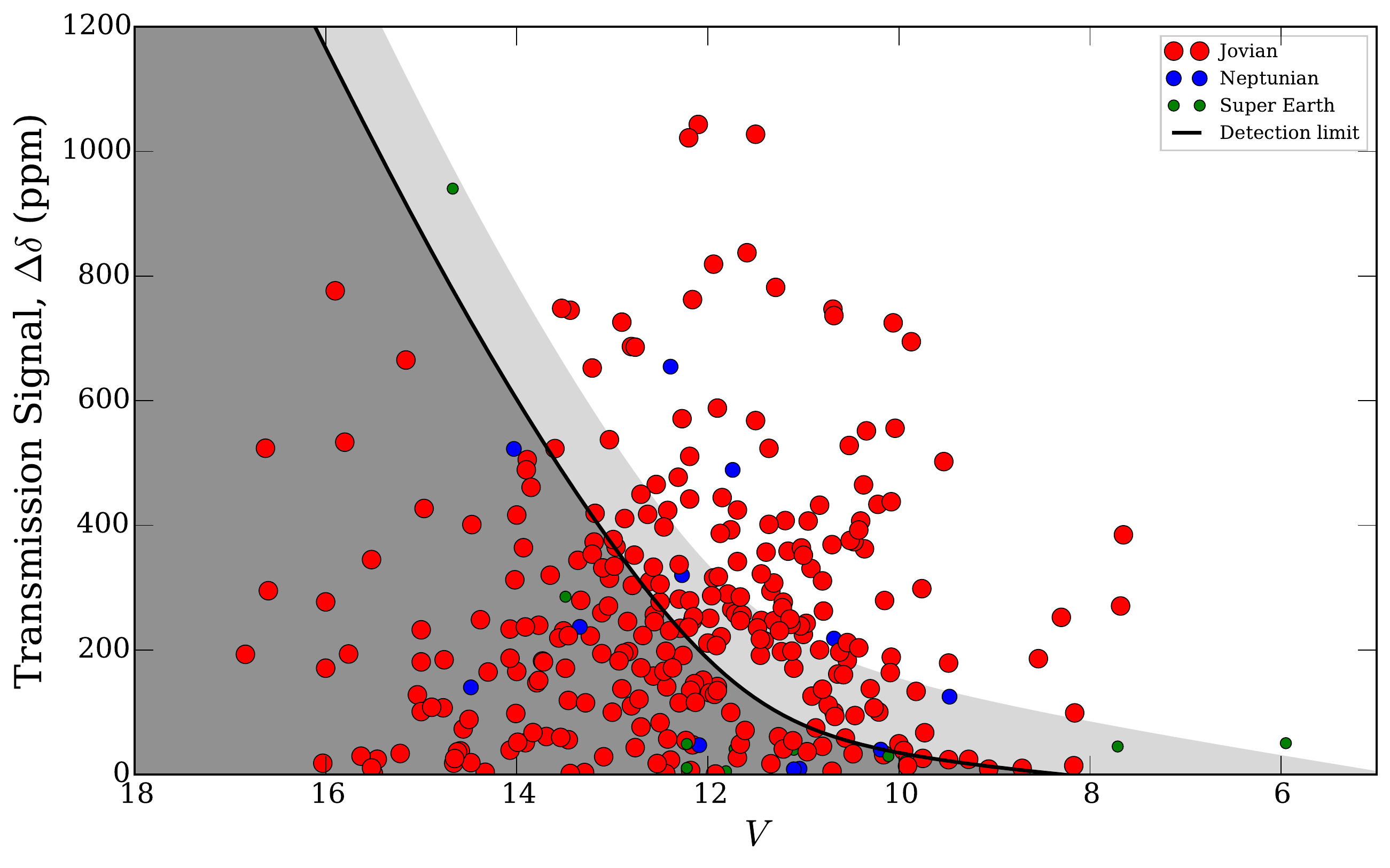}
   \end{tabular}
   \end{center}
   \caption[example] 
   { \label{fig:translim} 
   The transmission signal ($\Delta\delta$) of known transiting exoplanets versus the $V$ magnitude of the host star. The line represents the detection limit of FORS2, estimated from previous transmission spectroscopy observations with this instrument.  Atmospheric signals of planets in the dark grey region are beyond the reach of the instrument, whereas those in the white area should be detectable with FORS2.  Light grey represents an area where the detection would be tentative.}
   \end{figure}

As a planet crosses the disc of its host star, along the line of sight, the received
radiation is diminished as a function of the planet to star disc size ratios, {$\delta=(R_p/R_{\star})^2$}. During a planetary transit, some of the light from the host star goes through the planet's atmosphere before reaching us. When observed at different wavelengths, the transit depth, directly linked to the apparent planetary radius, may vary, providing constraints on the height of the atmosphere, the chemical composition and the existence of cloud layers (see Ref.~\citenum{Burrows14} for a recent review). This is due to the 
discrete absorptions by the atmospheric gases of the planet. The ability to measure the planet size precisely at the particular wavelengths corresponding to {those} previously mentioned absorptions (the  ``transmission spectrum'' or ``radius spectrum'') allows one to infer the presence of various molecules. Until now, water, sodium, potassium and hazes, as well as the presence of clouds, have been identified by this challenging technique. The signal associated with such variations is calculated as {$\Delta\delta$= $2A_H H Rp/R_{\star}^2$}, where $A_H$ is the
scale height number (typically between 1 and 10) and $H(= \frac{k_BT}{\mu_{mol}g})$ is the atmospheric scale height\footnote{$k_B$ is Boltzmann's constant; $T$ the {equilibrium} temperature; $\mu_{mol}$ the mean molecular weight; and $g$ is the surface gravity of the planet.}. As $H$ is much smaller than $R_p$ (typically by one or two orders of magnitude), this effect is clearly tiny -- much smaller than the faint dimming due to the transit itself -- and requires uttermost precision. 
It is also limited to the planets orbiting close to their host star, as in order to have $H$ reasonably large, one needs the temperature to be high enough.
In Fig.~\ref{fig:translim}, we show the corresponding value of $\Delta\delta$ as a function of the visual magnitude of the host star for all planets from TEPCat for which there exists an estimate of the temperature.

\begin{table}[htbp]
\caption{Ground-based transmission spectroscopic studies as of now.}
\centering
\label{Tab:TS}
\begin{tabular}{|lllccl|}
\hline
{\bf Instrument} & {\bf Telescope} & {\bf Target} & {\bf $\lambda$-range} & {\bf Bin size} & {\bf Reference}\\
& & & (nm) & (nm) & \\
\hline
DOLORES & TNG & HAT-P-1 & 525--760 & 60  & Montalto et al. 2015 [\citenum{2015ApJ...811...55M}]\\\hline
FORS2 & VLT & GJ 1214&  780--1\,000& 20 & Bean et al. 
2010 [\citenum{Bean10}]\\
& & GJ 1214& 610--850& 10,20 & Bean et al.  2011 [\citenum{2011ApJ...743...92B}]\\
 & & WASP-19& 560--820& 16--22.5 & Sedaghati et al.
2015 [\citenum{Sedaghati15}]\\
 & & WASP-17& 570--825& 5--20 & Sedaghati et al.
2016a [\citenum{Sedaghati16a}]\\
 & & WASP-19& 400--1\,000& 5--20 & Sedaghati et al.
2016b [\citenum{Sedaghati16}]\\
 & & WASP-49& 730--1\,000& 10 & Lendl et al. 2016 [\citenum{2016AA...587A..67L}]\\
 \hline
GMOS & Gemini & HAT-P-32 & 520--930& 14 & Gibson et al.
2013a [\citenum{2013MNRAS.436.2974G}]\\
{(N/S)} & &  WASP-29 & 515--720& 15  & Gibson et al.
2013b [\citenum{2013MNRAS.428.3680G}]\\
& & WASP-12 & 720--1\,010& 15 & Stevenson et al.
2014 [\citenum{2014AJ....147..161S}]\\ 
\hline
IMACS & Magellan & WASP-6 & 480--860& 20  &  Jord\`an et al.
2013 [\citenum{2013ApJ...778..184J}]\\
\hline
LDSS-3C & Magellan & HAT-P-26 & 720--1\,000 & 12.5 & Stevenson et al. 2016 [\citenum{2016ApJ...817..141S}]\\\hline
MMIRS & Magellan & WASP-19 & 1\,250--2\,350& 100 & Bean et al.
2013  [\citenum{2013ApJ...771..108B}]\\
\hline
MODS & LBT & HAT-P-32 & 330--1\,000& 11  & Mallonn \& Strassmeier 2016 [\citenum{2016AA...590A.100M}]\\\hline
MOSFIRE & Keck & GJ3470 & 1\,960--2\,390 & 40 & Crossfield et al. 2013 [\citenum{Crossfield13}]\\\hline
OSIRIS & GTC & HAT-P-19& 560--770 & 5--20 & Mallonn et al.
2015 [\citenum{Mallonn15}]\\
& & WASP-43& 540--920 & 10,25 &  Murgas et al.
2014 [\citenum{2014AA...563A..41M}]\\
& & HAT-P-32 &518--918& 20  & Nortmann et al. 2016 [\citenum{2016arXiv160406041N}]\\
 & & TrES-3 & 530--930 & 25 &
Parviainen et al. 2016 [\citenum{Parviainen16}]\\
\hline
\end{tabular}
\end{table}

To achieve the level of accuracy needed for such studies, while at the same time retaining enough spectral resolution, one needs to collect a large number of photons per wavelength bin. This can only be achieved with space telescopes -- Spitzer and the Hubble Space Telescope have produced most results -- or the largest ground-based telescopes.
Among the latest of the HST results is the comparative study\cite{2016Natur.529...59S} of ten hot Jupiters over the wavelength range from 300 nm to 5~$\mu$m that reveals a continuum from clear to cloudy atmospheres, with clouds and hazes being the cause of weaker spectral signatures. 

Because time on space telescopes is a scarce resource, it is important to ensure that ground-based facilities are also able to {fulfill} the necessary requirements\footnote{Here, we are only concerned with those instruments that have a wide spectral range and enough spectral resolution (i.e. we do not consider studies done with photometry, nor those looking at some specific lines only).}. Moreover, future spaced-based observatories will focus on the infrared, such that ground-based instrumentation will be the only way to probe the optical transmission spectra of exoplanets -- a crucial wavelength regime to understand the physics of exoplanet atmospheres, in particular to determine the mean molecular weight of the atmosphere and atmospheric scale height from measuring the Rayleigh scattering slope. 
Several results have been obtained in this respect, summarised\footnote{We hope this list is complete as of writing, but apologise in case we missed any study.} in Table~\ref{Tab:TS}, which clearly reveals that most large ground-based facilities have been used to perform transmission spectroscopy, but this has required -- or still requires -- understanding the systematic effects that plague such measurements. As an example, with GMOS, Huitson et al.\cite{2015AAS...22512405H} report photometric precisions per spectral bin of 200-600 ppm. 

At ESO, in the optical range, the only available instrument for this purpose is the FOcal Reducer and  low-dispersion Spectrograph (FORS2) attached to the 8.2-m Unit Telescope 1 (UT1). As seen in Table~\ref{Tab:TS}, after some pioneering work done by J. Bean and colleagues in 2010 and 2011, no other results based on this instrument appeared before our own study in 2015\cite{Sedaghati15}. We will come back to this later, after introducing in more detail the instrument.

\section{The FORS2 Instrument}

With {its} variety of modes, FORS2 is the Swiss army knife of ESO's Very Large Telescope\cite{Appenzeller98,Rupprecht10}. FORS2 is indeed capable {of} imaging, polarimetry, long-slit low and medium resolution spectroscopy and multi-object spectroscopy, using either movable slitlets or masks, in the wavelength range from 330 to 1\,100 nm. The last mode is particularly useful for transmission spectroscopy, as it can be used to do time-resolved spectroscopy of a planet-hosting star as well as some comparison stars for performing differential spectrophotometry. The field of view of FORS2 is $6.8\arcmin\times6.8\arcmin$ and the pixel size is 0.25\arcsec (when using the typical  2$\times$2 {binning}).

At the beginning of the VLT operations, there were two FORS instruments -- FORS1 and FORS2. In order to leave space for second generation instruments at the VLT, FORS1 was dismounted and stored in 2009, with some of its components being merged in a `new', hybrid FORS2.

 \begin{figure} [htbp]
   \begin{center}
   \begin{tabular}{c} 
   \includegraphics[width=17cm]{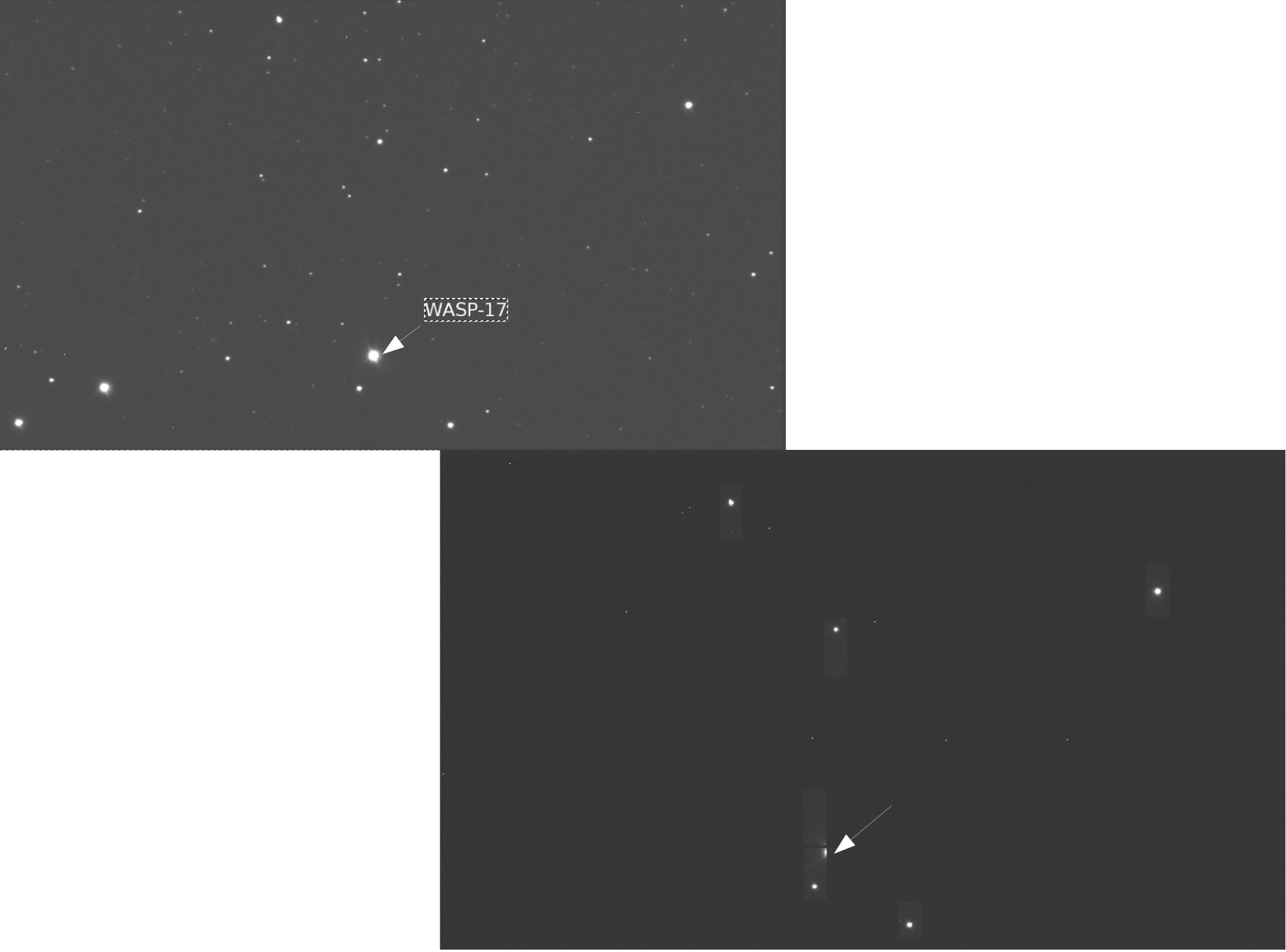}
   \end{tabular}
   \end{center}
   \caption[example] 
   { \label{Fig:bad} 
An unfortunate case where no useful data could be obtained. In this case, the instrument is not to blame, as it seems that the observer forgot to do pre-imaging, and possibly underestimated the proper motion of the target (indicated by the arrow), and so did not centre precisely enough the (wide) slits. The upper image is the acquisition image, while the lower one is the through-slit image.}
   \end{figure} 

 \begin{figure} [htbp]
   \begin{center}
   \begin{tabular}{cc} 
   \includegraphics[width=9.5cm]{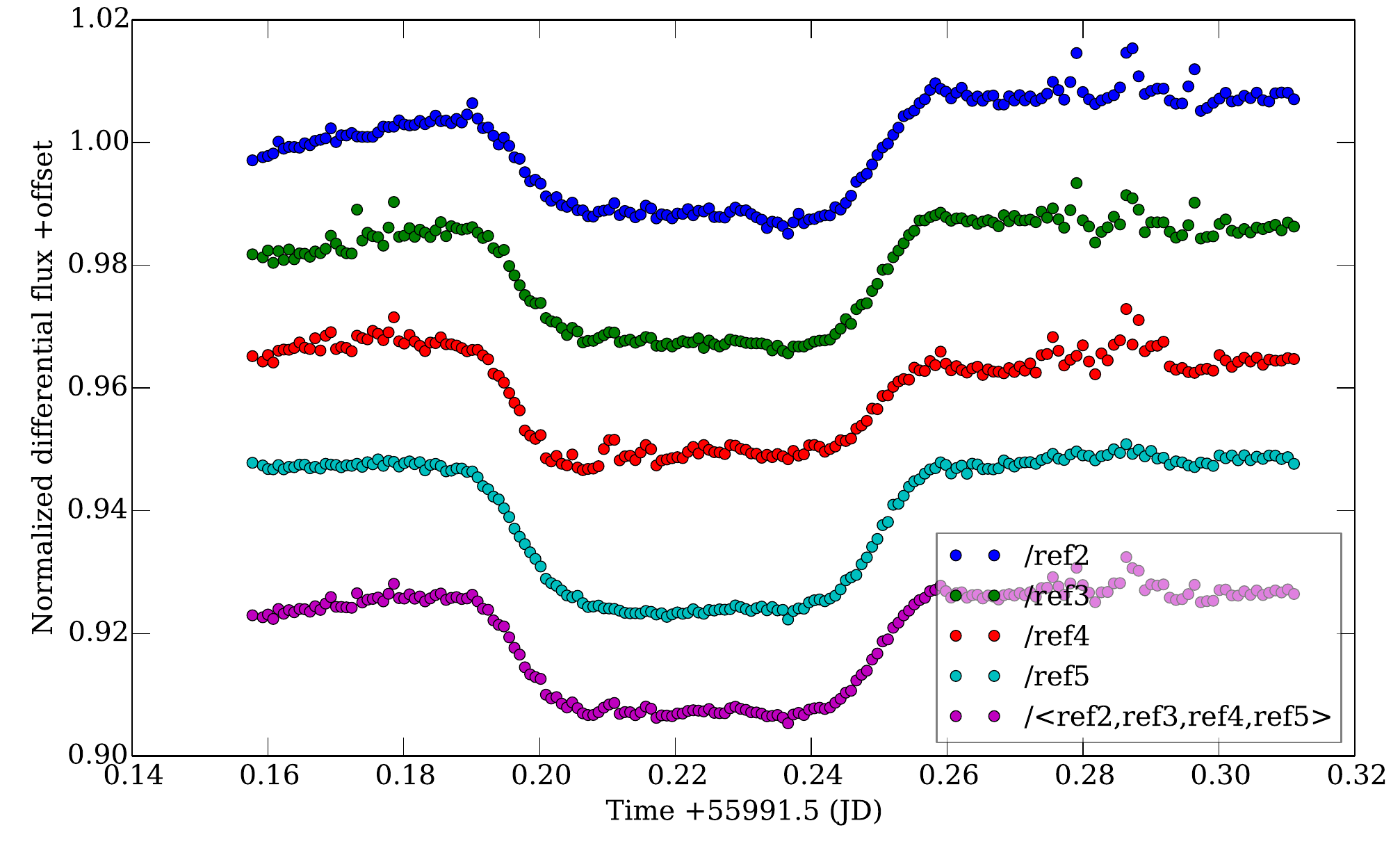}&\includegraphics[width=6.4cm]{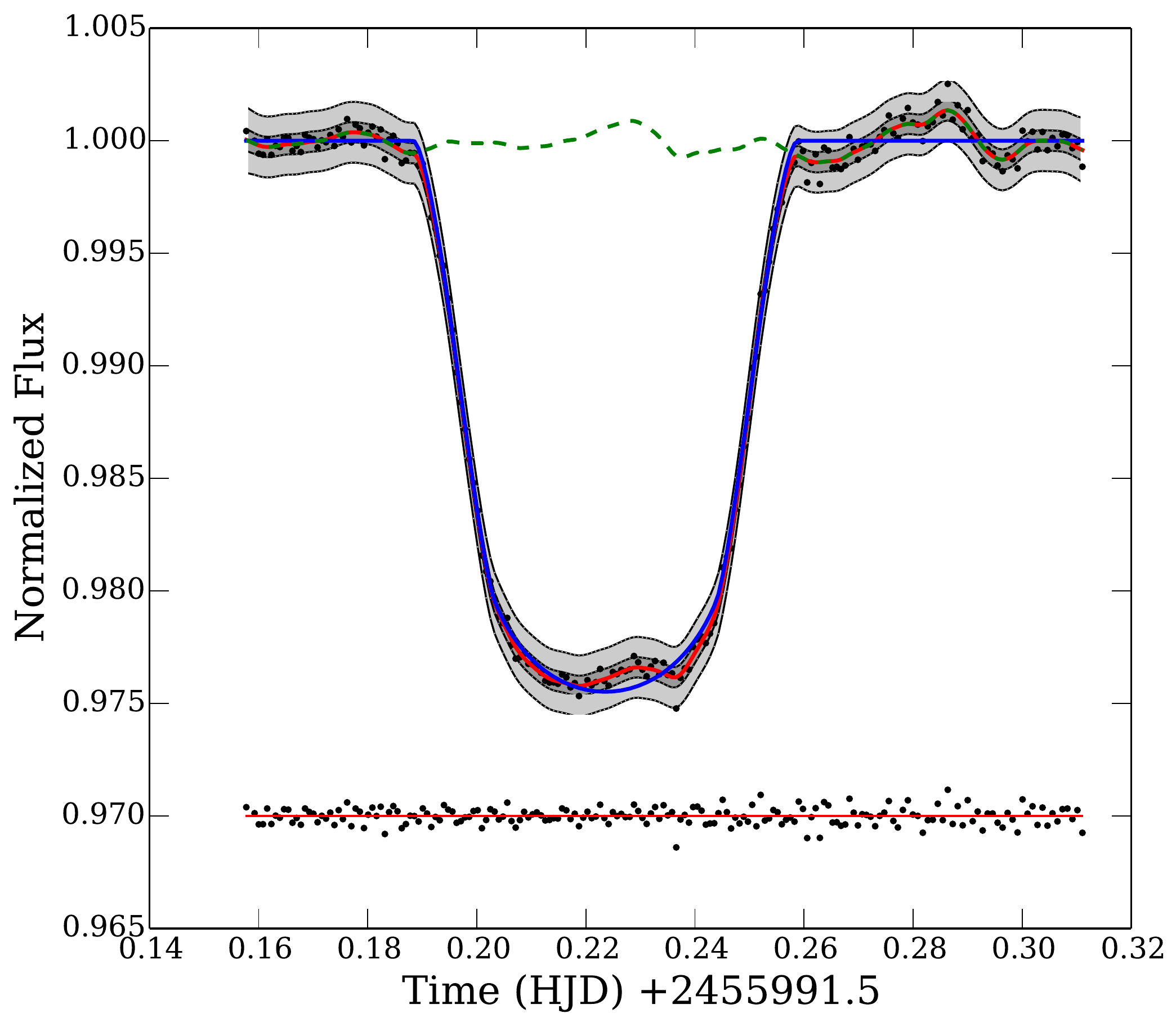}\\
   \end{tabular}
   \end{center}
   \caption[example] 
   { \label{Fig:pre} 
One transit obtained with FORS2 before the LADC prism exchange. On the left, we show the differential light curves of the target with respect to some comparison stars, as well as relative to their average. Large systematics are clearly present. This is then also visible in the final, detrended light curve shown on the right.}
   \end{figure}

 \begin{figure} [htbp]
   \begin{center}
   \begin{tabular}{c} 
   \includegraphics[width=13cm]{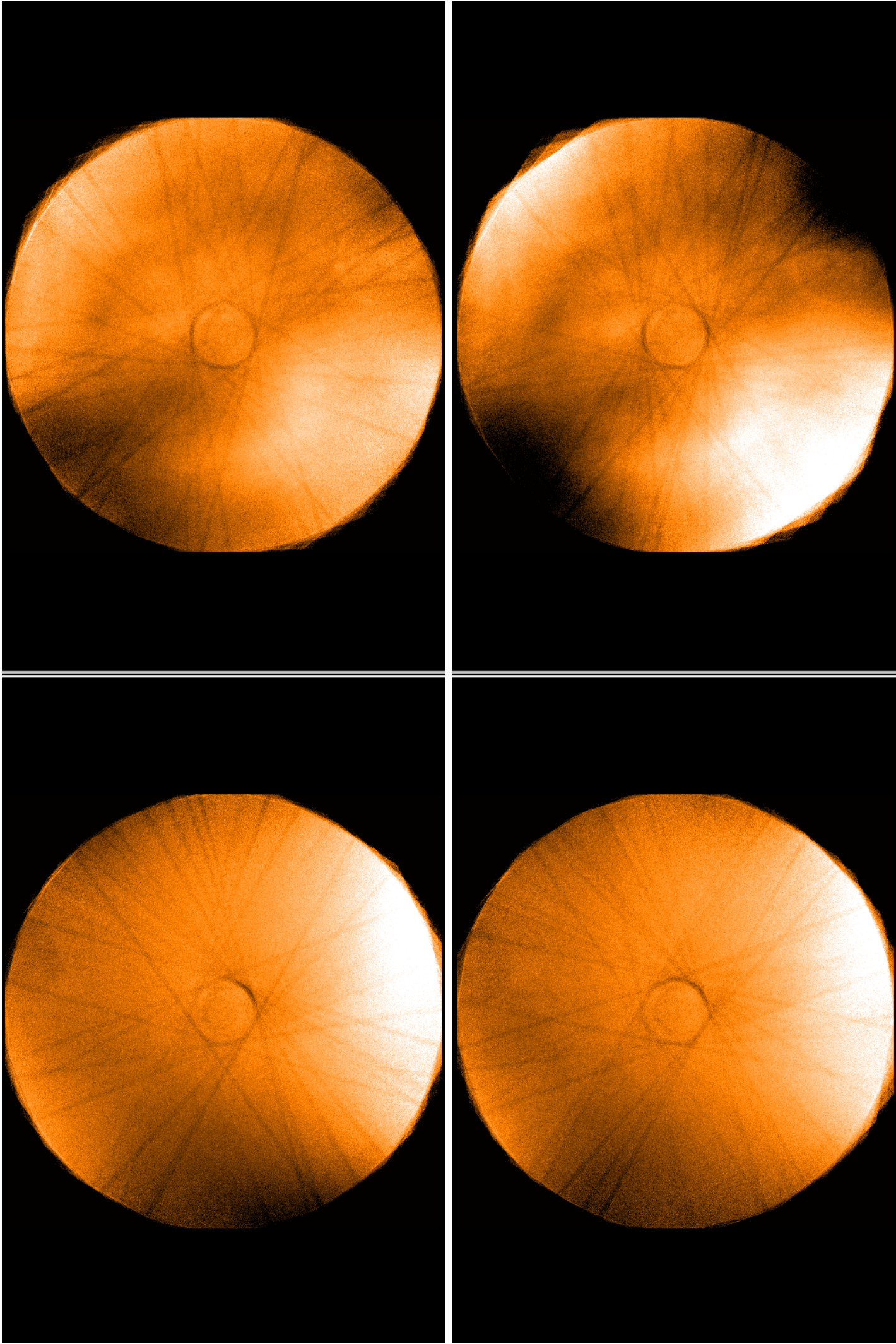}
   \end{tabular}
   \end{center}
   \caption[example] 
   { \label{Fig:sky flats} 
Comparison of rotator-angle dependent features seen in FORS2 twilight sky flats with the R\_SPECIAL (right) and I\_BESS (left) filters, before (top) and after (bottom) the LADC prism exchange. The false-colour scale represents artefacts between 0.995 (black) to 1.005 (white). Please see Ref.~\citenum{Boffin16} for details.}
   \end{figure} 
 
In front of the FORS instruments is a longitudinal atmospheric dispersion corrector (LADC), allowing it, unlike VIMOS for example, to perform multi-object spectroscopy (that requires a fixed rotator angle on sky, and can thus not be done at the parallactic angle) at high airmass. The design of the FORS2 LADC consists of two prisms of opposite orientation that are moved linearly with respect to each other, between 30 mm (park position) and 1100 mm. The forward prism does the dispersion correction, while the second prism corrects the pupil tilt, so that what remains is a variable image shift depending on the distance between the two prisms\cite{Avila97}.

Bean et al.\cite{Bean10} have shown the potential of FORS2 in producing transmission spectra for exoplanets even in the regime of mini-Neptune and super-Earth. They used FORS2 to obtain the transmission spectrum of GJ 1214b between wavelengths of 780 and 1\,000 nm, showing that the lack of features in this spectrum rules out cloud-free atmospheres composed primarily of hydrogen. This result was recently confirmed\cite{Kreidberg}. However, except for this ground-based pioneering result, all further attempts to use FORS2 for exoplanet transit studies have apparently failed. A questionnaire sent to the PIs of these programmes made it clear that in many cases the data obtained were affected by large systematics that prevented its use for transmission spectroscopy. 

We have looked into the FORS2 archive for these data and analysed them. Although in some cases, the fault may lie in the fact that this is a rather new technique and users needed to gain experience in developing the best strategy, or needed to better prepare the observations  (with the most extreme case being shown in Fig.~\ref{Fig:bad}), our {analyses} indeed reveal a high level of systematics as shown in Fig.~\ref{Fig:pre}.
Not all datasets seem to have been affected, though, as shown by the recent analysis\cite{2016AA...587A..67L} of a dataset obtained during this period.

Moehler et al.\cite{Moehler10} studied the twilight flat fields obtained with the two FORS instruments and found structures that rotated with the field rotator (see the top panels of Fig.~\ref{Fig:sky flats}) -- structures that had a significant impact on photometric measurements. These authors concluded that the origin of these structures was the LADC and, more precisely, the degradation of the antireflective coatings of the prisms that form the LADC.
It is these same structures that are 
most likely  the cause of the systematics seen in the transmission spectra, as also reported by Berta et al.\cite{Berta11}. 

To address this, the prisms of the FORS2 LADC were exchanged with the ones of FORS1, after having removed their coating, which was also degraded. A battery of tests {were} performed to ensure the exchanged prisms did not affect adversely the image quality of the instrument. The tests also confirmed that the LADC was still efficient at correcting the atmospheric dispersion up to an airmass of 1.6 (see Ref.~\citenum{Boffin15}, which explains in detail the tests and results and shows the resulting improvements). This led to a clear reduction of the systematics as seen in the FORS2 sky flats, collected over many months\cite{Boffin16}. As shown in Fig.~\ref{Fig:sky flats}, the small-scale structures clearly visible in the old data are gone, leaving only a gradient across the field.  

 \begin{figure} [htbp]
   \begin{center}
   \begin{tabular}{c} 
   \includegraphics[width=17cm]{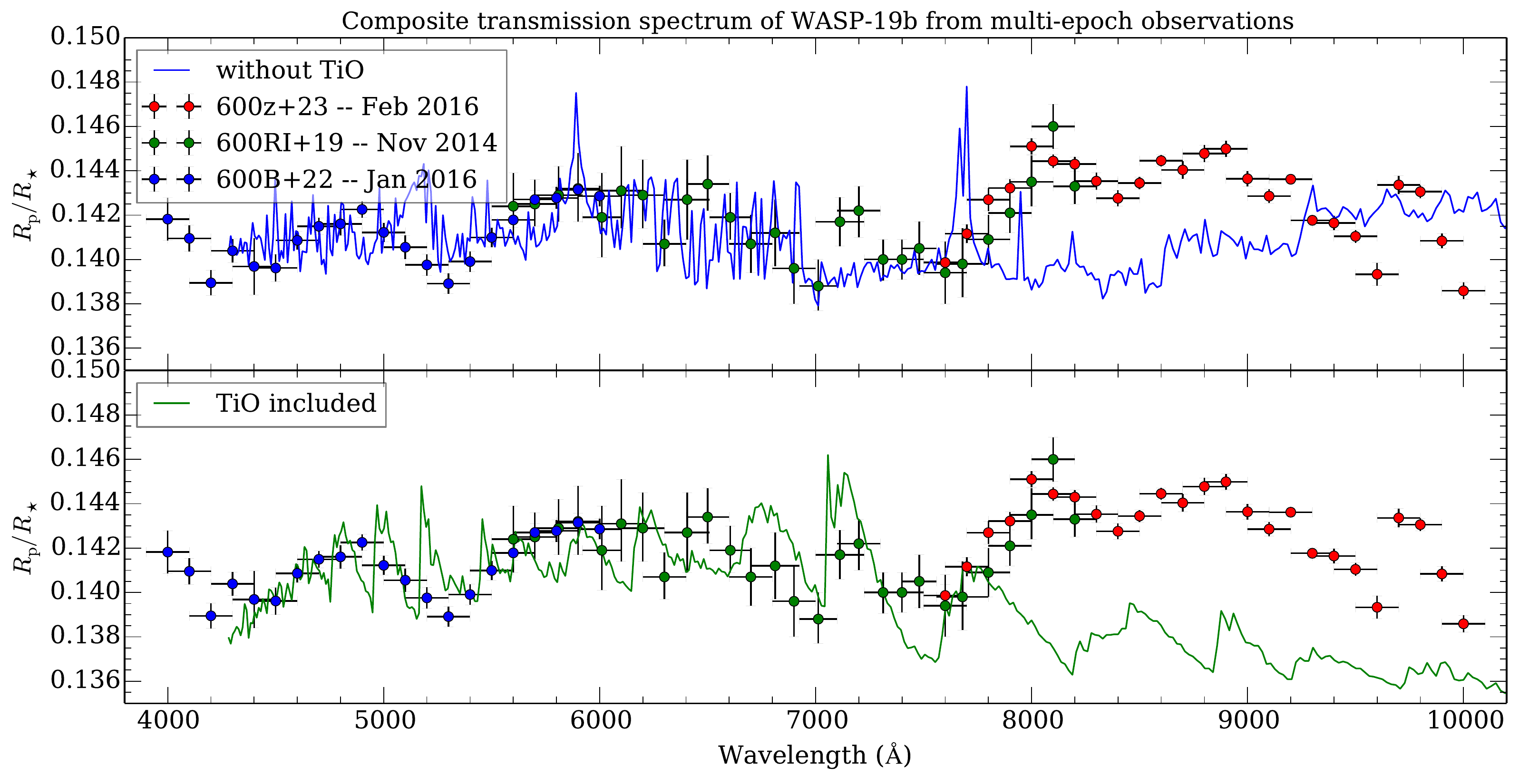}
   \end{tabular}
   \end{center}
   \caption[example] 
   { \label{fig:wasp19} 
Composite transmission spectrum of WASP-19b obtained with three FORS2 grisms at 3 separate epochs, compared to atmospheric models of solar composition, one with TiO molecules included and the other without (from Ref.~\citenum{Sedaghati15} and Sedaghati et al., in prep.). It is quite reassuring to see that where the grisms overlap in wavelength, the values for the planetary radius agree. The observations also reveal some unusual features redwards of $\sim$8\,000 \AA, which are as yet not explained by any model.}
   \end{figure} 

After the prism exchange, several transits of the hot Jupiter WASP-19b were observed and a transmission spectrum obtained over a wide wavelength range\cite{Sedaghati15,Sedaghati16,Sedaghati16a} (see Fig.~\ref{fig:wasp19}). The resulting spectrum, consistent over several transits and when using different grisms, indicate that the systematics that affected FORS2 have been significantly reduced and that residuals at the level of 200 ppm can now be obtained, an order of magnitude improvement compared to some of the datasets obtained before the prism exchange. These residuals can still be reduced depending on the target, the atmospheric conditions and an adequate observing strategy. One can thus hope that FORS2 is back in business for transmission spectroscopy and in the next and final section we provide hints for the best strategy to perform such measurements.

\section{Strategy for transmission spectroscopy with FORS2}
Based on our experience and tests, we have now been able to define the best possible strategy to perform transmission spectroscopy with FORS2. We assume that most of these points should also apply to other ground-based multi-object spectrographs. We encourage future users of FORS2 to follow these points when devising their observations.

\begin{itemize}
\item Put the target and reference stars as much as possible in the middle of the CCD. The wavelength coverage depends on the horizontal position of the object on the CCD. To guarantee the widest coverage and the best overlap between the target and the references, they should be placed in the middle of the CCD. This is easily done if there is only one reference star, by changing the position angle on sky. This fact should actually dictate, among others mentioned below, which reference stars are taken.
\item Use stars of similar magnitudes, as this allows a better comparison, without loss of S/N. This also ensures that the exposure time is not driven by one of the stars only, and therefore also guarantees that the flux of all stars will be similar. This is critical to mitigate any possible non-linearity effects of the CCD.
\item Use stars of similar colours, so as to have similar spectra to compare to, especially for the broad-band light curve\footnote{This shouldn't be so critical for the light curves in the narrow spectral channels, although it wouldn't harm if the stars have similar colours. Priority, however, should be given to have comparison stars as bright or brighter than the target.}. It is therefore useful to do pre-imaging of the field through several filters to establish the best possible comparison stars (this would also make sure that any high proper-motion of the target is correctly handled when making the masks), unless the colours of all stars are already available in catalogues.  
\item The maximum pixel intensity of the brightest object should have at least 40\,000 ADUs. The FORS2 CCDs are linear up to the saturation limit, 65\,000 ADUs, but it is safer not to go too close to this limit. 
\item Constantly check the counts on the raw spectra and, if needed, adjust the exposure time\footnote{Some observers are less keen to adjust the exposure time during the whole sequence, in order to have an homogeneous sampling. This, however, requires estimating carefully how the flux will change with airmass and should allow for improvement in image quality, without saturating the spectra.}, especially if there are significant changes in airmass.
\item Try not to change any parameters during the whole sequence (except, if needed, the exposure time -- see above), and make sure there is enough telescope tracking time for the whole sequence as well as enough number of exposures to cover the full duration.
\item Use a very long out-of-transit baseline, before and after the transit, and if possible do simultaneous photometry to monitor any stellar activity and the presence of stellar spots -- preferably in a filter matching the wavelength range of the spectroscopic observations.
\item If possible (that is, from a scheduling point of view and if the transit is short enough), try to avoid observing through the meridian, i.e. observe the full sequence before or after the meridian. Going through the meridian implies the largest field rotation and this seems to lead to additional systematics in some cases.
\item Use the {\tt 200 kHz, 2$\times$2, low} read-out mode as this will shorten the read-out time, and thanks to the low gain, allows increasing the exposure time before reaching the saturation limit. This thus {helps to reduce} the overheads and increase the time resolution of the observations, as well as increase the S/N obtained on a single spectrum. 
\item Observations should be done under clear conditions. If the atmosphere is not very stable, with thin (or thick!) clouds passing, it may be better to abort the observations and give back the telescope than to waste valuable telescope time for useless data.
\item Although this has not been characterised yet when used for transmission spectroscopy, for very bright targets it can be useful to consider using the ``Virtual Slit'', which uses the active optics of the telescope to spread the light in the $y-$direction (i.e. perpendicular to the wavelength dispersion) onto up to 6$\arcsec$ (that is, up to 24 binned pixels). This allows for longer integration times -- allowing to observe bright targets without too much dead-time, and is the almost equivalent to defocusing (but not similar, and in one direction only) that is used in photometry.
\item Use wide enough slits, i.e. equal or larger than 15$\arcsec$. This ensures that there are no slit losses, but also that in case there is a faint companion to your target or comparison stars, it also always stays in the slit. Obviously, this implies that the spectral resolution will then be determined by the image quality, typically 0.8--1$\arcsec$.
\item Use long enough slits to make sure to have enough sky to correct the spectra, but it is advisable to also put one or two slits on an empty sky (on both chips), to have a good estimate of the sky contamination for each exposure. This should also be as aligned as possible to the science targets. 
\item Put the LADC in park position (30 mm) and in simulation during the whole observation. This allows minimising any systematics coming from the LADC, i.e. any additional moving parts. This means, however, that the differential refraction is not corrected, but as the slits are long and wide, this should not imply slit losses.
\item Make sure to have also made a mask with narrow slits (0.4--1$\arcsec$), to be used for taking wavelength calibrations, as those done with the wide slits will be useless (e.g., won't be accepted by the FORS2 pipeline). Ask for more than one wavelength calibration frames (3 or 5).
\item Take enough spectroscopic flats (7 is the standard number from the calibration plan, but you may wish to ask for 20 or more) with both masks, if possible before and after the observations.  
\item Assess in real time the quality of the observations, and derive the broadband differential light curve as the observations progress\footnote{This requires that you have requested calibrations to be done {\bf prior} to your observations, which may not always be possible.}. This would give a clear indication if something needs to be changed in the experimental setup. 
\end{itemize}

Clear skies!

 

\bibliographystyle{spiebib} 

\end{document}